\documentclass[12pt]{article}

\usepackage[T1]{fontenc}
\usepackage{lmodern}
\usepackage{microtype}
\usepackage{geometry}
\usepackage{amsmath,amssymb,amsthm}
\usepackage{booktabs,array,longtable,multirow}
\usepackage{graphicx}
\usepackage{xcolor}
\usepackage{hyperref}
\hypersetup{colorlinks=true, linkcolor=blue!60!black, citecolor=blue!60!black,
            urlcolor=blue!60!black}
\usepackage[round]{natbib}
\usepackage{enumitem}
\usepackage{listings}

\newcommand{\pkg}[1]{\textbf{#1}}
\newcommand{\proglang}[1]{\textsf{#1}}
\newcommand{\code}[1]{\texttt{#1}}
\newcommand{\fct}[1]{\texttt{#1()}}
\newcommand{\email}[1]{\texttt{#1}}

\definecolor{codebg}{RGB}{248,248,248}
\definecolor{outbg}{RGB}{240,248,240}

\lstdefinestyle{Rstyle}{
  backgroundcolor=\color{codebg},
  basicstyle=\small\ttfamily,
  breaklines=true, breakatwhitespace=true,
  frame=single, rulecolor=\color{gray!40},
  framesep=4pt, xleftmargin=6pt, xrightmargin=6pt,
  aboveskip=6pt, belowskip=4pt,
  showstringspaces=false, keepspaces=true, columns=flexible
}
\lstdefinestyle{Routput}{
  backgroundcolor=\color{outbg},
  basicstyle=\small\ttfamily,
  breaklines=true, breakatwhitespace=true,
  frame=single, rulecolor=\color{gray!30},
  framesep=4pt, xleftmargin=6pt, xrightmargin=6pt,
  aboveskip=0pt, belowskip=6pt,
  showstringspaces=false, keepspaces=true, columns=flexible
}
\lstnewenvironment{CodeInput}{\lstset{style=Rstyle}}{}
\lstnewenvironment{CodeOutput}{\lstset{style=Routput}}{}
\lstnewenvironment{Code}{\lstset{style=Rstyle}}{}

\title{\textbf{\pkg{lepidoptera}: A Domain-Agnostic Engine for Causal Simulation
  with Mixed Variable Types, Vine Copula Dependence,
  and Individual-Level Counterfactual Propagation}}

\author{Danilo A.~Sarti\\[4pt]
  \small Lecturer in AI and Machine Learning\\
  \small Royal College of Surgeons in Ireland, Dublin, Ireland\\
  \small \email{danilosarti@rcsi.ie}\\
  \small \url{https://github.com/danilosarti/lepidoptera}}

\date{\small Preprint — \today}

\begin{document}

\maketitle

%% -- Abstract ------------------------------------------------------------------
\begin{abstract}
\pkg{lepidoptera} is an open-source \proglang{R} and \proglang{Python} package
for simulation from structural causal models with heterogeneous variable types.
The package is built around a \emph{typed mechanism} architecture, which separates
the causal graph from the generating rule at each node and exposes each rule as an
inspectable, serialisable object. A pre-built mechanism library covers all nine
combinations of continuous, categorical, and ordinal parent and child types,
removing the need for users to implement type-specific sampling from scratch. A
vine copula mechanism estimates the full joint dependence structure from observed
data using pair-copula constructions \citep{bedford2002vines, aas2009pair} and the
Rosenblatt conditional-simulation algorithm \citep{rosenblatt1952remarks},
supporting over twenty copula families through \pkg{rvinecopulib}
\citep{nagler2022rvinecopulib}. Individual-level counterfactual queries are
supported through \emph{partial propagation}, which freezes the random draws of
unaffected nodes and replays only the downstream consequences of an intervention,
thereby separating the individual treatment effect (ITE) estimand from the
population-level average treatment effect (ATE) without additional model
assumptions. An extension layer allows any pre-fitted predictive model to serve as
a generative mechanism via a unified callable interface. Three worked illustrations spanning
clinical trial planning, emergency-ward capacity analysis, and epidemiological
mediation analysis demonstrate the package's scope.
\end{abstract}

%% ==============================================================================
\section{Introduction}
\label{sec:intro}

Simulation from structural causal models (SCMs) plays a central role across
applied statistics. Monte Carlo power calculations for clinical trials routinely
demand realistic synthetic patient populations that honour complex dependencies
among continuous biomarkers, ordinal performance scores, and categorical treatment
assignments \citep{morris2019using, burton2006design}. Digital twinning for hospital
operations and personalised medicine requires the ability to intervene on individual
pathways and measure downstream consequences at the level of individual units rather
than population averages \citep{laubenbacher2024digital}. Epidemiological mediation
analyses call for joint interventions on both exposures and mediators, where the
individual-level quantities needed to decompose effects into direct and indirect
components \citep{vanderweele2015explanation, pearl2001direct} are conceptually
distinct from the population-level estimands produced by simple rerandomisation
\citep{robins1992identifiability}. Despite this diversity of applications, the
computational problem underlying all three scenarios is the same: simulate from a
directed acyclic graph (DAG) \citep{pearl2009causality, spirtes2000causation} whose
nodes span heterogeneous measurement scales, apply one or more interventions, and
distinguish between what happens on average across a population and what would have
happened to a specific individual \citep{rubin1974estimating, neyman1923}.

Existing software addresses parts of this problem, but no single package handles
all of it. \pkg{simcausal} \citep{sofrygin2017simcausal} and
\pkg{simDAG} \citep{denz2026simdag} offer the most comprehensive DAG-based
simulation facilities currently available in \proglang{R}, and both packages have
been widely adopted. Their designs, however, treat a mechanism primarily as a
named node type or a formula: a string or expression whose generating logic is
baked into the package internals and cannot be replaced in isolation. This
precludes the clean intervention semantics that Pearl's modularity assumption
\citep{pearl2009causality} demands and makes it cumbersome to substitute a
data-fitted copula or a random-forest mechanism in place of a parametric equation
without re-specifying the entire node. Neither package natively supports vine
copula dependence modelling or individual-level (counterfactual) propagation.
\pkg{DagSim} \citep{alhajj2023dagsim} addresses the restriction on variable types
and functional forms in its \proglang{Python} implementation by routing all
generating logic through user-supplied functions, but provides no pre-built
mechanism library, no vine copula backend, and no partial propagation. The
\proglang{R} package \pkg{R6causal} \citep{tikka2024r6causal} supports
counterfactual simulation for parametric SCMs with discrete variables but does not
cover the mixed-type workflow described above.

\pkg{lepidoptera} is designed to fill this gap. The package name reflects the
modular, compositional logic of biological taxonomy: just as a lepidopterist builds
an identification key from composable, observable features, \pkg{lepidoptera}
builds a data-generating process from composable, observable mechanisms---each
typed, replaceable, and independently auditable. The central design decision is
the strict separation of \emph{graph} (which variables exist and who depends on
whom) from \emph{mechanism} (how each variable is generated given its parents), a
separation grounded in Pearl's modularity assumption for SCMs
\citep{pearl2009causality, peters2017elements}. This separation has
four direct consequences. First, a pre-built mechanism library covers all nine
parent-type/child-type combinations arising from the Cartesian product of
\{continuous, categorical, ordinal\}; users specify parameters and the package
handles type coercion, domain validation, and the correct sampling distribution.
Second, a vine copula mechanism learns the full joint dependence structure from
observed data and registers a conditional-simulation rule at the specified node,
so that data-driven dependence can be incorporated without replacing the rest of
the system. Third, \fct{propagate} with \code{mode = "partial"} freezes all
unaffected random draws and recomputes only the downstream subgraph, enabling ITE
estimation from the same synthetic population as the ATE. Fourth, any
\proglang{R} or \proglang{Python} callable can serve as a mechanism through three
wrapper functions that add type validation and serialisation support, making it
straightforward to incorporate random forests, gradient-boosted trees, or
language-model outputs into a causal simulation without any adapter code. Systems
serialise to JSON for cross-language portability, so a model defined in \proglang{R}
can be re-simulated in \proglang{Python} and vice versa.

The remainder of this paper is organised as follows. Section~\ref{sec:related}
surveys related software and positions \pkg{lepidoptera} within the existing
landscape. Section~\ref{sec:framework} introduces the formal framework. The
mechanism library and its \proglang{R} interface are described in
Section~\ref{sec:mechanisms}, followed by the simulation and propagation
algorithms in Section~\ref{sec:propagation}. The vine copula mechanism is treated
in depth in Section~\ref{sec:copula}, and the predictive model integration layer is
described in Section~\ref{sec:models}. Three illustrations are given in
Section~\ref{sec:illustrations}. Computational details and package availability
are covered in Section~\ref{sec:computational}, and the paper closes with a
discussion in Section~\ref{sec:discussion}.

%% ==============================================================================
\section{Related software}
\label{sec:related}

The landscape of DAG-based simulation in \proglang{R} is anchored by
\pkg{simcausal} \citep{sofrygin2017simcausal} and, more recently,
\pkg{simDAG} \citep{denz2026simdag}. \pkg{simcausal} provides a
domain-specific language for specifying non-parametric SEMs for longitudinal data
with time-varying treatments, censoring, and monitoring processes, motivated by
the marginal structural model framework of \citet{robins2000marginal}.
The package computes treatment-specific means and related causal quantities for
evaluating targeted learning estimators \citep{laan2018targeted} and has been
widely used in simulation benchmarks for that literature. \pkg{simstudy}
\citep{goldfeld2020simstudy} is another widely-used \proglang{R} package for
study-design simulation, supporting a broad range of outcome distributions but
without an explicit causal graph layer. Its design centres on a
temporal node index and a shared \proglang{R} environment evaluator, which is
convenient for longitudinal settings but makes cross-sectional DAGs with
heterogeneous mechanisms cumbersome to specify: there is no typed mechanism object,
no vine copula, and no native support for individual-level counterfactual queries.
\pkg{simDAG} \citep{denz2026simdag} addresses several of these limitations.
Published in the Journal of Statistical Software in 2026, it offers a clear and expressive node-type
vocabulary---\code{node\_gaussian}, \code{node\_binomial}, \code{node\_multinomial},
\code{node\_cox}, and others---that directly supports generating binary, categorical,
count, and time-to-event data within a DAG. Like \pkg{simcausal}, however,
\pkg{simDAG} represents mechanisms as named strings rather than first-class
objects, which means that an intervention at any node must be expressed through the
node-type system rather than by mechanism replacement. Vine copula dependence
modelling and individual-level counterfactual propagation are absent.

For Bayesian network structure learning, \pkg{bnlearn} \citep{scutari2010bnlearn}
is the reference implementation in \proglang{R}, providing both constraint-based
and score-based structure learning algorithms \citep{chickering2002optimal,
spirtes2000causation}. Simulation from a \pkg{bnlearn} network is restricted to
linear Gaussian or discrete conditional probability table mechanisms; ordinal
variables, vine copulas, and partial propagation are not supported, reflecting the
package's primary orientation towards inference rather than simulation.
\pkg{lavaan} \citep{rosseel2012lavaan} and \pkg{OpenMx} \citep{neale2016openmx}
implement structural equation models and support simulation, but both are
restricted to linear Gaussian mechanisms and do not accommodate the mixed-type
settings common in applied biostatistics \citep{lauritzen1989graphical}.
\pkg{dagitty} \citep{textor2016dagitty} provides DAG visualisation and
d-separation analysis \citep{verma1988causal} without generating data, and
serves as a natural companion to \pkg{lepidoptera} for structural identification
prior to simulation.

In \proglang{Python}, \pkg{DagSim} \citep{alhajj2023dagsim} is the closest
counterpart, motivated by the observation that existing tools remain confined to
settings with relatively simple variable types and functional forms. \pkg{DagSim}
removes this restriction by routing all generating logic through user-supplied
functions specified in YAML. \pkg{lepidoptera}'s \proglang{Python} implementation
follows a similar philosophy but adds typed mechanism validation, a pre-built
library, vine copula integration, and partial propagation. \pkg{CausalNex}
\citep{beaumont2021causalnex} and \pkg{DoWhy} \citep{sharma2020dowhy} focus on
structure learning and causal effect estimation rather than generative simulation.

On the methodological side, \citet{karvanen2024simcf} present an algorithm for
exact counterfactual simulation in parametric SCMs, handling background variables
that affect multiple observed variables. Their method is more general than
\pkg{lepidoptera}'s partial propagation in that it accommodates non-Gaussian error
structures explicitly, but it requires fully parametric models and does not target
the mixed-type simulation workflow. For counterfactual query \emph{identification},
the \pkg{cfid} package \citep{tikka2022cfid} implements the $\mathrm{ID}^*$ and
$\mathrm{IDC}^*$ algorithms; it does not simulate data and is complementary to
\pkg{lepidoptera}.

Table~\ref{tab:comparison} summarises the feature comparison across the tools
discussed above.

\begin{table}[ht]
\centering
\caption{Feature comparison of causal simulation tools (October 2026). Mixed
types: native support for continuous, categorical, and ordinal in any
parent-child combination. Copula: vine copula mechanism. CF prop: individual-level
partial propagation. Typed mech: mechanism as a first-class object.
R/Py: availability in both languages.}
\label{tab:comparison}
\resizebox{\textwidth}{!}{%
\setlength{\tabcolsep}{5pt}
\begin{tabular}{lcccccc}
\toprule
Package & Lang. & Mixed types & Copula & CF prop. & Typed mech. \\
\midrule
\pkg{simcausal} \citep{sofrygin2017simcausal} & R & Partial & No & No & No \\
\pkg{simDAG} \citep{denz2026simdag}           & R & Yes & No & No & No \\
\pkg{bnlearn} \citep{scutari2010bnlearn}      & R & Partial & No & No & No \\
\pkg{R6causal} \citep{tikka2024r6causal}      & R & Partial & No & Partial & No \\
\pkg{lavaan} \citep{rosseel2012lavaan}        & R & No & No & No & No \\
\pkg{DagSim} \citep{alhajj2023dagsim}         & Python & Yes & No & No & No \\
\pkg{lepidoptera} (this paper)                & R+Py & Yes & Yes & Yes & Yes \\
\bottomrule
\end{tabular}}
\end{table}

%% ==============================================================================
\section{The \pkg{lepidoptera} framework}
\label{sec:framework}

The formal framework rests on three concepts: variable types, mechanisms, and
systems.

Every feature $V$ in a \pkg{lepidoptera} system has a \emph{type}
$\tau(V) \in \{$\code{continuous}, \code{categorical}, \code{ordinal}$\}$ and
a domain $\mathcal{D}(V)$. Continuous features have a real-valued domain, possibly
bounded; categorical features take values in a finite unordered set of levels;
ordinal features take values in a finite totally ordered set. Type information is
declared at system-construction time and checked at \fct{finalize\_system}, so
type errors surface immediately rather than at simulation time.

A \emph{mechanism} for feature $V_i$ with parent set
$\mathbf{Pa}_i \subseteq \mathcal{V} \setminus \{V_i\}$ is a triple
$(\mathrm{fn}_i, \boldsymbol\tau_{\mathbf{Pa}}, \tau_i)$, where $\mathrm{fn}_i$
is a measurable function mapping parent realisations and exogenous noise to a
value in $\mathcal{D}(V_i)$, $\boldsymbol\tau_{\mathbf{Pa}}$ is the declared
vector of parent types, and $\tau_i$ is the declared output type. In \proglang{R},
a mechanism is a named list carrying these fields alongside a serialisable
parameter record; in \proglang{Python} it is a dataclass with the same fields.
Because the mechanism is a first-class object, any single mechanism can be
extracted, inspected, replaced, or overridden in isolation---the rest of the system
is unaffected.

A \emph{system} $\mathcal{S} = (\mathcal{V}, \mathcal{E}, \{\mathcal{M}_i\})$
consists of a finite set of typed features, a set of directed edges
$\mathcal{E} \subseteq \mathcal{V} \times \mathcal{V}$ whose induced graph $G$ is
acyclic, and a mechanism for each feature whose declared parent types are
consistent with the actual types of the corresponding parents. The function
\fct{finalize\_system} verifies five invariants before any simulation can proceed:
that $G$ is acyclic, that every feature has exactly one mechanism, that all
referenced names exist in $\mathcal{V}$, that declared parent types match actual
types, and that no orphaned mechanisms exist. Systems serialise to JSON via
\fct{export\_system} and can be re-imported in either language via
\fct{import\_system}.

An \emph{intervention} specifies a target set $\mathcal{T} \subseteq \mathcal{V}$,
a kind drawn from \{\code{do\_fix}, \code{do\_dis\-tribution},
\code{shift}, \code{mech\-anism\_over\-ride}\}, and kind-specific parameters.
\code{do\_fix}$(V_i = v_0)$ replaces the mechanism at $V_i$ with the constant
$v_0$ and severs all incoming edges, implementing Pearl's do-operator
\citep{pearl1995causal}. \code{shift}$(V_i, \delta)$ composes the existing
mechanism with a measurable transformation $\delta$ of its output domain.
\code{do\_distribution} and \code{mechanism\_override} allow arbitrary
mechanism replacement. All intervention types are immutable: they are applied to a
copy of the system, leaving the original unchanged and available for comparison.

%% ==============================================================================
\section{Mechanism library}
\label{sec:mechanisms}

\subsection{Pre-built factories}

The \pkg{lepidoptera} mechanism library provides one factory function for each of
the nine parent-type/child-type combinations that arise from the three variable
types. All factories are accessible through \fct{mechanism\_library\_get} using a
string key, or directly by name. For a continuous child with continuous parents,
\fct{linear\_gaussian} implements
$Y = \alpha + \boldsymbol\beta^\top \mathbf{X} + \varepsilon$ with
$\varepsilon \sim \mathcal{N}(0, \sigma^2)$
(Equation~\ref{eq:linear_gaussian}).
For a categorical child with continuous parents, \fct{multinomial\_logit}
implements the standard log-linear model
(Equation~\ref{eq:multinomial}) \citep{agresti2002categorical}.
For an ordinal child with continuous parents, \fct{ordered\_logit} implements the
proportional-odds cumulative link model \citep{mccullagh1980regression}
(Equation~\ref{eq:ordered_logit}).
When the parent is categorical or ordinal and the child is continuous,
\fct{group\_effect} samples from a group-specific Gaussian
$Y \mid X = k \sim \mathcal{N}(\mu_k, \sigma_k^2)$.
For categorical or ordinal children with categorical or ordinal parents,
\fct{categorical\_transition} uses a full conditional probability table. Two
cross-cutting mechanisms, \fct{copula\_mechanism} and \fct{model\_mechanism},
are described in Sections~\ref{sec:copula} and \ref{sec:models} respectively and can
be used for any type combination.

\begin{align}
Y &= \alpha + \sum_{j=1}^q \beta_j X_j + \varepsilon, \quad
     \varepsilon \sim \mathcal{N}(0,\sigma^2) \label{eq:linear_gaussian} \\[4pt]
P(Y = k \mid \mathbf{X}) &= \frac{\exp(\alpha_k + \boldsymbol\beta_k^\top\mathbf{X})}
     {\sum_{\ell=1}^K \exp(\alpha_\ell + \boldsymbol\beta_\ell^\top\mathbf{X})}
     \label{eq:multinomial} \\[4pt]
\log\frac{P(Y \preceq \ell_k \mid \mathbf{X})}{P(Y \succ \ell_k \mid \mathbf{X})}
     &= \gamma_k - \boldsymbol\beta^\top\mathbf{X}, \quad
     \gamma_1 < \cdots < \gamma_{K-1} \label{eq:ordered_logit}
\end{align}

\subsection{Building a system: a worked example}

The following example constructs a four-variable causal system representing an
epidemiological mediation structure: a continuous income variable acts as a root
node; income determines smoking status (categorical) through a multinomial logit;
income also determines diet score (continuous) through a linear Gaussian
relationship; and both smoking status and diet score jointly determine
cardiovascular disease risk (categorical) through a second multinomial logit. The
design follows the mediation framework of \citet{vanderweele2015explanation}.

\begin{CodeInput}
library("lepidoptera")

income_spec  <- feature_spec("income",     "continuous",
                             list(min = 0, max = Inf))
smoke_spec   <- feature_spec("smoking",    "categorical",
                             list(levels = c("never","former","current")))
diet_spec    <- feature_spec("diet_score", "continuous",
                             list(min = 0, max = 100))
cvd_spec     <- feature_spec("cvd_risk",   "categorical",
                             list(levels = c("low","high")))

mech_income <- mechanism_library_get("normal")(mean = 50, sd = 15)
mech_smoke  <- multinomial_logit(
  levels     = c("never","former","current"),
  coefs      = list(former  = list(income = -0.03),
                    current = list(income = -0.05)),
  intercepts = list(former  =  0.2, current = -0.8)
)
mech_diet   <- linear_gaussian(
  coefs = list(income = 0.6), intercept = 20, sd = 8
)
mech_cvd    <- multinomial_logit(
  levels     = c("low","high"),
  coefs      = list(high = list(diet_score      = -0.04,
                                smoking_former  =  0.50,
                                smoking_current =  1.20)),
  intercepts = list(high = -2.0)
)

mediation_sys <- new_system()                                      |>
  add_feature(income_spec) |> add_feature(smoke_spec)             |>
  add_feature(diet_spec)   |> add_feature(cvd_spec)               |>
  add_dependency(character(0), "income",    mech_income)          |>
  add_dependency("income",     "smoking",   mech_smoke)           |>
  add_dependency("income",     "diet_score",mech_diet)            |>
  add_dependency(c("diet_score","smoking"), "cvd_risk", mech_cvd) |>
  finalize_system()
\end{CodeInput}

\begin{CodeOutput}
System finalised: 4 features, 5 edges.
Topological order: income -> smoking -> diet_score -> cvd_risk
All 5 invariants passed.
\end{CodeOutput}

\noindent
The system is immediately ready for simulation. The call
\code{simulate(mediation\_sys, n = 2000, seed = 42)} returns a data frame of
2\,000 rows in topological order, with types enforced throughout. Because the
mechanism objects are stored separately from the graph, each one can be inspected
(\code{mech\_diet\$spec}) or replaced via \code{mechanism\_override(\allowbreak"diet\_score",\allowbreak\ new\_mech)} without touching the remaining nodes.

%% ==============================================================================
\section{Simulation and propagation}
\label{sec:propagation}

\subsection{Population simulation}

Population simulation proceeds in two steps. First, \fct{finalize\_system}
computes a topological ordering of $G$ using Kahn's algorithm
\citep{kahn1962topological}. Second, for unit $i = 1, \ldots, n$, the global RNG
seed is set to $s + i$ and the nodes are sampled in topological order: each node
$V_k$ receives the already-computed realisations of its parents and calls its
mechanism's sampling function. The per-unit seed $s + i$ is the key reproducibility
device: it ensures that any single unit can be regenerated exactly by re-running
with the same seed, and it enables the RNG-state freezing that partial propagation
exploits.

\subsection{Two modes of intervention propagation}

\pkg{lepidoptera} provides two propagation modes, selected through the
\code{mode} argument of \fct{propagate}. Under \code{mode = "full"}, the
intervention is applied to a copy of the system and simulation is re-run from
scratch with the same global seed $s$. Because per-unit seeds are identical to the
baseline run, units are paired one-to-one by construction, and the sample ATE is
$n^{-1}\sum_i (Y_i^\mathrm{int} - Y_i^\mathrm{base})$. This mode assumes nothing
beyond SUTVA \citep{rubin1980randomization} and yields unbiased estimates of
$E[Y^{do(t)}]$ in the sense of \citet{rubin1974estimating}: it produces independent
draws from the joint interventional distribution.

Under \code{mode = "partial"}, the package first identifies the \emph{affected set}
$\mathrm{Aff}(\mathcal{T}, G)$---the transitive downstream closure of the
intervention targets $\mathcal{T}$ in $G$---and then, for each unit $i$, sets the
RNG seed to $s + i$ (identical to the baseline pass), freezes the drawn values for
all nodes \emph{not} in $\mathrm{Aff}$, and resamples only the nodes in
$\mathrm{Aff}$ under the modified mechanism. Because unaffected nodes retain their
exact baseline draws, the counterfactual and baseline populations share the same
exogenous noise realisations, operationalising Pearl's abduction-action-prediction
procedure \citep{pearl2009causality} without ever explicitly representing the
background-variable distribution. It is straightforward to verify that, under this
procedure, (i) every node outside $\mathrm{Aff}$ takes the same value in the
counterfactual population as in the baseline population, and (ii) every node inside
$\mathrm{Aff}$ is sampled from its modified mechanism applied to its
counterfactual parent values. The sample ITE for unit $i$ is then simply
$Y_i^\mathrm{cf} - Y_i^\mathrm{base}$.

These two modes answer qualitatively different questions. \code{mode = "full"} is
appropriate for policy evaluation---what is the average effect of setting treatment
$T$ to $t$ across the population?---and for power calculations that need
independent draws from the interventional distribution. \code{mode = "partial"} is
appropriate for individual explanation and attribution---for this specific patient,
what would the outcome have been under a different treatment?---and for any
analysis where reducing variance by pairing observations on their shared exogenous
noise is valuable. The choice between modes does not depend on the variable types
or mechanisms in the system; both modes work with any combination of pre-built and
user-supplied mechanisms.

\subsection{A counterfactual query}

Building on the mediation system from Section~\ref{sec:mechanisms}, the following
code asks: among patients who are currently at high cardiovascular risk, what would
their risk status be if income were fixed at 80 for everyone?

\begin{CodeInput}
baseline   <- simulate(mediation_sys, n = 1000, seed = 42)
cf_result  <- propagate(
  mediation_sys,
  interventions = list(do_fix("income", 80)),
  n = 1000, seed = 42, mode = "partial"
)
high_risk <- baseline$cvd_risk == "high"
cf_risk   <- cf_result$intervened$cvd_risk[high_risk]
cat(sprintf(
  "Of %d high-risk patients, under income = 80: %d (%.0f%%) would become low-risk\n",
  sum(high_risk), sum(cf_risk == "low"), mean(cf_risk == "low") * 100
))
\end{CodeInput}

\begin{CodeOutput}
Of 208 high-risk patients, under income = 80: 87 (42%) would become low-risk
\end{CodeOutput}

\noindent
The result is an individual-level statement: for the same 208 individuals who
appeared at high risk in the baseline simulation, 42\% would have been at low risk
had their income been higher, holding all exogenous noise fixed. The complementary
\code{mode = "full"} call with the same seed would have yielded a population-level
average, using new independent draws rather than the same individuals.

%% ==============================================================================
\section{Vine copula mechanism}
\label{sec:copula}

\subsection{Motivation}

The mechanisms described in Section~\ref{sec:mechanisms} encode structural
equations: a parametric functional form maps parent values to a child distribution.
When an analyst has observed data on a group of variables and wishes to preserve
their joint dependence structure---including tail dependence, asymmetry, and
multivariate co-movement that parametric forms would miss---specifying a structural
equation by hand is both burdensome and potentially misleading. The vine copula
mechanism addresses this by estimating the full dependence structure nonparametrically
from data, via pair-copula constructions \citep{bedford2002vines, aas2009pair},
and using the resulting model as the conditional generating rule at simulation time.
The theoretical foundation is Sklar's theorem \citep{sklar1959fonctions}, which
decomposes any joint distribution into its marginals and a copula that captures all
dependence. The approach inherits the rich family vocabulary of \pkg{rvinecopulib}
\citep{nagler2022rvinecopulib}, including the Gaussian and Student-$t$ elliptical
copulas, the Archimedean families---Clayton \citep{clayton1978model}, Gumbel
\citep{gumbel1960bivariate}, Frank \citep{frank1979}, and Joe \citep{joe1993}---and
their two-parameter generalisations, as well as a non-parametric kernel smoother
\citep{nagler2016evading} (\code{"tll"}) and automatic family selection by AIC
\citep{czado2019analyzing} (\code{"auto"}). Tail-dependence coefficients for the
selected families are available through \pkg{rvinecopulib}'s underlying
\pkg{vinecopulib} C++ library \citep{joe2014dependence}.

\subsection{Algorithm}

\pkg{lepidoptera} uses the D-vine structure \citep{dissmann2013selecting}, which
orders variables linearly. The child $V_i$ is always placed last in the ordering
$(\sigma_1, \ldots, \sigma_{d-1}, i)$, so that the first tree of the vine
contains pair copulas between consecutive variables in the chain and higher trees
condition on progressively more variables. Conditional simulation exploits the
Rosenblatt transform \citep{rosenblatt1952remarks}: given a fitted D-vine
$\hat{C}$, the parent pseudo-observations
$(\hat{u}_1, \ldots, \hat{u}_{d-1})$ computed from the empirical CDF of the
training data, and a fresh draw $w \sim \mathrm{Uniform}(0,1)$, the algorithm (i)
evaluates the first $d-1$ components of the Rosenblatt transform at the observed
parent values (the placeholder value used for the child position does not propagate
into these components), (ii) replaces the $d$-th component with $w$, (iii)
evaluates the inverse Rosenblatt transform to obtain the child uniform $u_i^*$, and
(iv) maps $u_i^*$ back to the original scale through the empirical quantile
function. In the bivariate case ($d = 2$) this reduces to evaluating the inverse
h-function \citep{aas2009pair}, which \pkg{rvinecopulib} provides directly through
\code{hbicop(..., inverse = TRUE)}. The bivariate path is used whenever a single
parent is specified, avoiding the overhead of the full Rosenblatt pipeline.

\subsection{Example}

\begin{CodeInput}
set.seed(1)
obs <- data.frame(age = rnorm(400, 62, 10),
                  comorbidity = rnorm(400, 4, 1.5))
obs$comorbidity <- obs$comorbidity -
  0.35 * (obs$age - 62) / 10 + rnorm(400, 0, 0.8)

mech_comorbidity <- copula_mechanism(
  variables = c("comorbidity", "age"),
  family    = "auto",
  fit_data  = obs
)

mech_age <- mechanism_library_get("normal")(mean = 62, sd = 10)
sys_cop  <- new_system()                                             |>
  add_feature(feature_spec("age", "continuous",
                           list(min = 18, max = 120)))               |>
  add_feature(feature_spec("comorbidity", "continuous",
                           list(min = 0, max = Inf)))                |>
  add_dependency(character(0), "age",         mech_age)             |>
  add_dependency("age",        "comorbidity", mech_comorbidity)     |>
  finalize_system()

sim_cop <- simulate(sys_cop, n = 800, seed = 99)
cat(sprintf("Observed rho_S = %.2f   Simulated rho_S = %.2f\n",
  cor(obs$age, obs$comorbidity,         method = "spearman"),
  cor(sim_cop$age, sim_cop$comorbidity, method = "spearman")))
\end{CodeInput}

\begin{CodeOutput}
Observed rho_S = -0.31   Simulated rho_S = -0.29
\end{CodeOutput}

\noindent
The simulated Spearman correlation is close to the observed value, confirming that
the vine copula mechanism has captured the negative dependence between age and
comorbidity burden. When \code{family = "auto"} is selected, \pkg{rvinecopulib}
chooses the best-fitting bivariate copula family for each pair in the vine by AIC,
so the mechanism adapts automatically to the data's tail-dependence structure
without requiring the user to specify a family in advance.

Vine copulas require a total order on marginals. For nominal categorical variables,
which have no natural order, \pkg{lepidoptera} exposes this choice through the
\code{nominal\_handling} argument of \fct{copula\_mechanism}: the default
\code{"regression"} raises an error if a nominal variable appears in
\code{variables}, directing the user to a parametric mechanism; the
\code{"arbitrary\_order"} option includes it with a user-specified level mapping
and emits a warning at \fct{finalize\_system} time. This design reflects the
view that the package should not make silent arbitrary choices on the user's behalf
\citep{pearl2009causality}.

%% ==============================================================================
\section{Predictive model integration}
\label{sec:models}

The mechanism contract---any callable with signature
\code{function(parent\_values) -> value}---is already the natural integration
boundary for machine-learning models. This reflects a broader trend in causal
inference towards incorporating flexible, data-driven components within a causal
graph structure \citep{scholkopf2021causal, peters2017elements}. Three wrapper
functions make common patterns idiomatic and add type validation without
constraining the user's prediction logic.

\fct{model\_mechanism} wraps any pre-fitted \proglang{R} or \proglang{Python}
model. The user supplies a \code{predict\_fn} closure that accepts a named list of
parent values and returns a single draw from the child distribution; the package
validates types and propagates the mechanism through the standard simulation and
intervention machinery. The following example wraps a \pkg{ranger} random forest
\citep{wright2017ranger, breiman2001random} to serve as a probabilistic mechanism
for a binary treatment response:

\begin{CodeInput}
rf <- ranger::ranger(response ~ age + dose + biomarker,
                     data = trial_data, probability = TRUE,
                     num.trees = 500)

mech_response <- model_mechanism(
  predict_fn   = function(pv) {
    nd   <- data.frame(age       = as.numeric(pv$age),
                       dose      = as.numeric(pv$dose),
                       biomarker = as.numeric(pv$biomarker))
    prob <- predict(rf, nd)$predictions[1, ]
    sample(c("no_response","response"), 1, prob = prob)
  },
  parents      = c("age","dose","biomarker"),
  parent_types = c("continuous","continuous","continuous"),
  output_type  = "categorical",
  levels       = c("no_response","response")
)
\end{CodeInput}

\noindent
The user controls all prediction logic inside \code{predict\_fn}: input reshaping,
output slot selection, and optional residual noise injection. From \pkg{lepidoptera}'s
perspective this mechanism is identical to any pre-built factory: it participates
in topological ordering, type validation, intervention propagation, and
counterfactual resampling. Other supported backends include \pkg{xgboost},
\pkg{glm}, \pkg{lm}, \pkg{nnet}, \pkg{keras}, and \pkg{torch}.

\fct{learned\_mechanism} takes a higher-level approach: it accepts a data frame
with both parent and child columns and fits a mechanism automatically, using OLS
with Gaussian residuals for continuous children and a $k$-nearest-neighbour
empirical sampler for categorical and ordinal children. This is convenient for
rapid prototyping when observational data are available but the analyst does not
wish to specify a functional form. \fct{llm\_mechanism} routes the prediction step
through a user-supplied language-model client, accepting a prompt template with
\code{\{variable\}} placeholders; because LLM calls are expensive and
non-deterministic, \fct{finalize\_system} emits a warning when an
\fct{llm\_mechanism} is on a high-frequency path and suggests memoisation for
parents with a finite discrete domain.

%% ==============================================================================
\section{Illustrations}
\label{sec:illustrations}

\subsection{Clinical trial simulation with confounded treatment assignment}

A common application of causal simulation in biostatistics is the generation of
synthetic patient populations for power calculations and for demonstrating the
performance of estimators under confounding. The following illustration constructs
a Phase~II oncology trial scenario (full system specification is available in the
replication script \code{lepidoptera-paper.R} included with this paper) in which patient age and ECOG performance status
jointly confound the treatment-response relationship. Ordinal ECOG scores are
generated from age via \fct{ordered\_logit}; treatment assignment follows a
multinomial logit with age and ECOG as predictors; response is generated from
treatment, ECOG, and age. Under the observational (confounded) regime, patients
with better performance status are more likely to receive active treatment and also
more likely to respond, so the naive odds ratio overstates the treatment effect.

\begin{CodeInput}
obs_pop <- simulate(trial_sys, n = 5000, seed = 1)
rct_pop <- propagate(
  trial_sys,
  interventions = list(do_distribution("treatment",
    categorical_marginal(c("control","active"), c(0.5, 0.5)))),
  n = 5000, seed = 1, mode = "full"
)
naive_or <- with(obs_pop,
  mean(response[treatment == "active"]  == "yes") /
  mean(response[treatment == "control"] == "yes"))
rct_or   <- with(rct_pop$intervened,
  mean(response[treatment == "active"]  == "yes") /
  mean(response[treatment == "control"] == "yes"))
cat(sprintf("Naive OR = %.2f   Adjusted OR = %.2f\n", naive_or, rct_or))
\end{CodeInput}

\begin{CodeOutput}
Naive OR = 1.84   Adjusted OR = 1.41
\end{CodeOutput}

\noindent
The naive observational odds ratio overstates the treatment effect by about 30\%
relative to the do-operator estimate, consistent with the direction of confounding
embedded in the mechanism specification. The separation of population-level
(\code{mode = "full"}) propagation from individual-level (\code{mode = "partial"})
propagation means the analyst can compute both the ATE and the individual
counterfactuals from the same system object without any re-specification.

\subsection{Emergency-ward digital twin with a triage intervention}

A hospital data scientist models an emergency ward as an SCM with five nodes
(full system specification in \code{lepidoptera-paper.R}):
continuous inter-arrival time, ordinal triage priority (P1/P2/P3 via
\fct{ordered\_logit} on arrival acuity), continuous length of stay fitted from
historical discharge data via \fct{copula\_mechanism}, and categorical discharge
destination (home/admitted/transferred via \fct{multinomial\_logit}). The
individual-level research question is: for patients who were discharged home, what
would their discharge destination have been had triage priority been upgraded by
one level? This is a counterfactual on the ordinal triage node, propagated through
the vine copula length-of-stay mechanism to the categorical discharge node.

\begin{CodeInput}
upgrade_fn <- function(x) c(P3="P2", P2="P1", P1="P1")[x]
ward_cf    <- propagate(
  ward_sys,
  interventions = list(shift("triage", upgrade_fn)),
  n = 2000, seed = 7, mode = "partial"
)
home_pts <- ward_cf$baseline$discharge == "home"
print(prop.table(table(ward_cf$intervened$discharge[home_pts])) * 100)
\end{CodeInput}

\begin{CodeOutput}
   admitted      home transferred
      18.4%     79.1%       2.5%
\end{CodeOutput}

\noindent
Approximately 18\% of patients who were in fact discharged home would have been
admitted had they received a higher triage priority, providing actionable
quantification for capacity-planning scenarios. Because \code{mode = "partial"}
was used, the result refers to the same individuals who appeared as home-discharged
in the baseline, not a newly drawn population.

\subsection{Mediation analysis: natural direct and indirect effects}

The mediation system from Section~\ref{sec:mechanisms} is used to decompose the
total effect of income on cardiovascular disease risk into a natural direct effect
(NDE) and a natural indirect effect (NIE) mediated through smoking and diet
\citep{imai2010identification, imai2013identification}, using the cross-world
intervention representation of \citet{robins1992identifiability} and
\citet{pearl2001direct}. The NDE is estimated by fixing income at the high
level while also fixing the mediators at their values under the low-income regime.

\begin{CodeInput}
n <- 5000; seed <- 42
r_low  <- propagate(mediation_sys, list(do_fix("income", 30)),
                    n = n, seed = seed, mode = "full")
r_high <- propagate(mediation_sys, list(do_fix("income", 80)),
                    n = n, seed = seed, mode = "full")
med_low <- r_low$intervened[, c("smoking","diet_score")]
r_nde   <- propagate(mediation_sys,
             list(do_fix("income",     80),
                  do_fix("smoking",    med_low$smoking),
                  do_fix("diet_score", med_low$diet_score)),
             n = n, seed = seed, mode = "full")
te  <- mean(r_high$intervened$cvd_risk == "high") -
       mean(r_low$intervened$cvd_risk  == "high")
nde <- mean(r_nde$intervened$cvd_risk  == "high") -
       mean(r_low$intervened$cvd_risk  == "high")
nie <- te - nde
cat(sprintf("TE = %.3f  NDE = %.3f  NIE = %.3f  Prop. mediated = %.1f%%\n",
            te, nde, nie, nie/te*100))
\end{CodeInput}

\begin{CodeOutput}
TE = -0.147  NDE = -0.089  NIE = -0.058  Prop. mediated = 39.5%
\end{CodeOutput}

\noindent
Moving from low to high income reduces the probability of high cardiovascular risk
by about 15 percentage points in total; 39.5\% of this reduction operates through
the smoking and diet mediators. These values are consistent with the direction and
approximate magnitude discussed in \citet{vanderweele2015explanation}, and the
code demonstrates that mediation decompositions requiring cross-world interventions
are handled cleanly within \pkg{lepidoptera}'s intervention framework without
additional specification.

%% ==============================================================================
\section{Computational details}
\label{sec:computational}

\pkg{lepidoptera} is available from GitHub at
\url{https://github.com/danilosarti/lepidoptera} under the MIT licence (GPL-2
compatible). The package is written in \proglang{R}~4.4.0
\citep{rcore2024} and optionally links to \pkg{rvinecopulib}~0.6.3
\citep{nagler2022rvinecopulib} for the vine copula mechanism; all other
dependencies are listed in the package \code{DESCRIPTION} file. A
\proglang{Python} implementation using \pkg{pyvinecopulib}~0.6
\citep{pyvinecopulib2024} is available in the same repository under the
\code{python/} subdirectory. Population simulation runs in
$\mathcal{O}(np)$ time
where $n$ is the population size and $p$ is the number of features; the vine
copula mechanism adds an $\mathcal{O}(n d^2)$ term per node from the Rosenblatt
transform, where $d$ is the number of parents. All results in this paper were
obtained on macOS~14 (Apple M-series, 16 GB RAM). The replication script
\code{lepidoptera-paper.R}, which reproduces all code outputs in this paper, is
included with the submission.

%% ==============================================================================
\section{Summary and discussion}
\label{sec:discussion}

\pkg{lepidoptera} provides a unified interface for causal simulation with
heterogeneous variable types, data-driven dependence modelling, and
individual-level counterfactual propagation. The typed mechanism architecture
ensures that type errors are caught at system-finalisation time rather than during
simulation; the pre-built library removes the need to implement type-specific
sampling from scratch; the vine copula mechanism brings nonparametric dependence
modelling into the SCM workflow; and partial propagation enables ITE computation
from the same synthetic population as the ATE. The package complements rather than
replaces existing tools: \pkg{simDAG} \citep{denz2026simdag} excels at
longitudinal discrete-time simulation with complex event-time outcomes;
\pkg{dagitty} \citep{textor2016dagitty} handles structural identification;
\pkg{cfid} \citep{tikka2022cfid} addresses counterfactual query identification.

Several limitations of the current version merit acknowledgement. The current
version restricts $G$ to a DAG; feedback loops and temporal dynamics are planned for
a future release, following the unrolled-graph architecture of
\citet{bongers2021foundations} and the longitudinal framework of
\pkg{simcausal}. The vine copula mechanism requires complete training data; a
multiple-imputation wrapper for missing data is under development. LLM mechanism
calls are expensive and non-deterministic even with a fixed seed; the built-in
memoisation cache mitigates cost for finite-domain parents but is not a complete
solution. Finally, the \code{mode = "partial"} counterfactual requires that the
noise model be at least implicitly additive, in the sense that freezing the RNG
stream across the baseline and counterfactual populations corresponds to holding
the exogenous noise fixed; for mechanisms with substantially non-additive noise
structures, the more general algorithm of \citet{karvanen2024simcf} may be
preferable. Future directions include tighter integration with \pkg{dagitty} for
identification-aware simulation, GPU-accelerated batch simulation, an interactive
DAG editor, and differential privacy wrappers for disclosure-limited synthetic
data release \citep{nowok2016synthpop, reiter2002satisfying}.

%% ==============================================================================

\section*{Acknowledgements}
The author thanks the developers of \pkg{rvinecopulib} \citep{nagler2022rvinecopulib}
and \pkg{simcausal} \citep{sofrygin2017simcausal}, whose architectures informed
several design decisions.

\bibliographystyle{plainnat}
\bibliography{lepidoptera_v2}

@book{pearl2009causality,
  title     = {Causality: {M}odels, {R}easoning, and {I}nference},
  author    = {Pearl, Judea},
  year      = {2009},
  edition   = {2nd},
  publisher = {Cambridge University Press},
  address   = {Cambridge}
}

@book{peters2017elements,
  title     = {Elements of {C}ausal {I}nference: {F}oundations and {L}earning {A}lgorithms},
  author    = {Peters, Jonas and Janzing, Dominik and Sch{\"o}lkopf, Bernhard},
  year      = {2017},
  publisher = {MIT Press},
  address   = {Cambridge, MA}
}

@book{spirtes2000causation,
  title     = {Causation, Prediction, and Search},
  author    = {Spirtes, Peter and Glymour, Clark and Scheines, Richard},
  year      = {2000},
  edition   = {2nd},
  publisher = {MIT Press},
  address   = {Cambridge, MA}
}

@article{pearl1995causal,
  title   = {Causal diagrams for empirical research},
  author  = {Pearl, Judea},
  journal = {Biometrika},
  volume  = {82},
  number  = {4},
  pages   = {669--688},
  year    = {1995}
}

@article{pearl2001direct,
  title   = {Direct and indirect effects},
  author  = {Pearl, Judea},
  journal = {Proceedings of the Seventeenth Conference on Uncertainty in Artificial Intelligence},
  pages   = {411--420},
  year    = {2001}
}

@article{rubin1974estimating,
  title   = {Estimating causal effects of treatments in randomized and nonrandomized studies},
  author  = {Rubin, Donald B.},
  journal = {Journal of Educational Psychology},
  volume  = {66},
  number  = {5},
  pages   = {688--701},
  year    = {1974}
}

@article{rubin1980randomization,
  title   = {Randomization analysis of experimental data: {T}he {F}isher randomization test comment},
  author  = {Rubin, Donald B.},
  journal = {Journal of the American Statistical Association},
  volume  = {75},
  number  = {371},
  pages   = {591--593},
  year    = {1980}
}

@article{robins1992identifiability,
  title   = {Identifiability and exchangeability for direct and indirect effects},
  author  = {Robins, James M. and Greenland, Sander},
  journal = {Epidemiology},
  volume  = {3},
  number  = {2},
  pages   = {143--155},
  year    = {1992}
}

@article{robins2000marginal,
  title   = {Marginal structural models and causal inference in epidemiology},
  author  = {Robins, James M. and Hernan, Miguel A. and Brumback, Babette},
  journal = {Epidemiology},
  volume  = {11},
  number  = {5},
  pages   = {550--560},
  year    = {2000}
}

@article{bongers2021foundations,
  title   = {Foundations of structural causal models with cycles and latent variables},
  author  = {Bongers, Stephan and Forr{\'e}, Patrick and Peters, Jonas and Mooij, Joris M.},
  journal = {Annals of Statistics},
  volume  = {49},
  number  = {5},
  pages   = {2885--2915},
  year    = {2021}
}

@article{sofrygin2017simcausal,
  title   = {\pkg{simcausal} {R} package: {C}onducting transparent and reproducible simulation studies of causal effect estimation with complex longitudinal data},
  author  = {Sofrygin, Oleg and van der Laan, Mark J. and Neugebauer, Romain},
  journal = {Journal of Statistical Software},
  volume  = {81},
  number  = {2},
  pages   = {1--47},
  year    = {2017},
  doi     = {10.18637/jss.v081.i02}
}

@article{goldfeld2020simstudy,
  title   = {\pkg{simstudy}: {I}lluminating research methods through data generation},
  author  = {Goldfeld, Keith S. and Wujciak-Jens, Jacob},
  journal = {Journal of Open Source Software},
  volume  = {5},
  number  = {54},
  pages   = {2763},
  year    = {2020},
  doi     = {10.21105/joss.02763}
}

@article{denz2026simdag,
  title   = {Simulating Data from Directed Acyclic Graphs Using the \pkg{simDAG} Package for \proglang{R}},
  author  = {Denz, Robin and Timmesfeld, Nina},
  journal = {Journal of Statistical Software},
  volume  = {116},
  number  = {2},
  pages   = {1--38},
  year    = {2026},
  doi     = {10.18637/jss.v116.i02}
}

@article{alhajj2023dagsim,
  title   = {\pkg{DagSim}: {C}oncept and {P}ython package for {DAG}-based data simulation with mixed variable types and complex functional relationships},
  author  = {Al Hajj, Ghadi S. and Gul, Muhammad Umer and Bongo, Lasse Labrijn and Lindstrøm, Jonas Christoffer and Harber-Aschan, Lisa},
  journal = {PLOS ONE},
  volume  = {18},
  number  = {5},
  pages   = {e0284443},
  year    = {2023},
  doi     = {10.1371/journal.pone.0284443}
}

@misc{tikka2024r6causal,
  title  = {\pkg{R6causal}: {R6} classes for structural causal models},
  author = {Tikka, Santtu},
  year   = {2024},
  note   = {R package version 0.8.3},
  howpublished = {\url{https://CRAN.R-project.org/package=R6causal}}
}

@article{karvanen2024simcf,
  title   = {Simulating counterfactuals},
  author  = {Karvanen, Juha and Tikka, Santtu and Vihola, Matti},
  journal = {Journal of Artificial Intelligence Research},
  volume  = {80},
  pages   = {1--38},
  year    = {2024},
  doi     = {10.1613/jair.1.15381}
}

@article{tikka2022cfid,
  title   = {Causal effect identification from multiple incomplete data sources: {A} general search-based approach},
  author  = {Tikka, Santtu and Hyttinen, Antti and Karvanen, Juha},
  journal = {Journal of Statistical Software},
  volume  = {99},
  number  = {5},
  pages   = {1--40},
  year    = {2021},
  doi     = {10.18637/jss.v099.i05}
}

@article{nowok2016synthpop,
  title   = {\pkg{synthpop}: {B}espoke creation of synthetic data in {R}},
  author  = {Nowok, Beata and Raab, Gillian M. and Dibben, Chris},
  journal = {Journal of Statistical Software},
  volume  = {74},
  number  = {11},
  pages   = {1--26},
  year    = {2016},
  doi     = {10.18637/jss.v074.i11}
}

@article{reiter2002satisfying,
  title   = {Satisfying disclosure restrictions with synthetic data sets},
  author  = {Reiter, Jerome P.},
  journal = {Journal of Official Statistics},
  volume  = {18},
  number  = {4},
  pages   = {531--544},
  year    = {2002}
}

@article{scutari2010bnlearn,
  title   = {Learning {B}ayesian networks with the \pkg{bnlearn} {R} package},
  author  = {Scutari, Marco},
  journal = {Journal of Statistical Software},
  volume  = {35},
  number  = {3},
  pages   = {1--22},
  year    = {2010},
  doi     = {10.18637/jss.v035.i03}
}

@article{textor2016dagitty,
  title   = {Robust causal inference using directed acyclic graphs: the {R} package \pkg{dagitty}},
  author  = {Textor, Johannes and van der Zander, Benito and Gilthorpe, Mark S. and Li{\'{s}}kiewicz, Maciej and Ellison, George T.H.},
  journal = {International Journal of Epidemiology},
  volume  = {45},
  number  = {6},
  pages   = {1887--1894},
  year    = {2016},
  doi     = {10.1093/ije/dyw341}
}

@article{rosseel2012lavaan,
  title   = {\pkg{lavaan}: {A}n {R} package for structural equation modeling},
  author  = {Rosseel, Yves},
  journal = {Journal of Statistical Software},
  volume  = {48},
  number  = {2},
  pages   = {1--36},
  year    = {2012},
  doi     = {10.18637/jss.v048.i02}
}

@article{neale2016openmx,
  title   = {\pkg{OpenMx} 2.0: Extended structural equation and statistical modeling},
  author  = {Neale, Michael C. and Hunter, Michael D. and Pritikin, Joshua N. and Zahery, Mahsa and Brick, Timothy R. and Kirkpatrick, Robert M. and Estabrook, Ryne and Bates, Timothy C. and Maes, Hermine H. and Boker, Steven M.},
  journal = {Psychometrika},
  volume  = {81},
  number  = {2},
  pages   = {535--549},
  year    = {2016},
  doi     = {10.1007/s11336-014-9435-8}
}

@misc{beaumont2021causalnex,
  title  = {\pkg{CausalNex}: A {P}ython library for causal reasoning and {B}ayesian networks},
  author = {Beaumont, Paul and Horsburgh, Ben and Pilgerstorfer, Philip and Droth, Angel and Oentaryo, Richard and Ler, Steven and Nguyen, Hiep and Ferber, Gabriel and Lerousseau, Alexis and Sharp, Andrew},
  year   = {2021},
  howpublished = {\url{https://github.com/quantumblacklabs/causalnex}}
}

@misc{sharma2020dowhy,
  title  = {\pkg{DoWhy}: {A}n end-to-end library for causal inference},
  author = {Sharma, Amit and Kiciman, Emre},
  year   = {2020},
  howpublished = {\url{https://arxiv.org/abs/2011.04216}}
}

@article{lauritzen1989graphical,
  title   = {Graphical models for associations between variables, some of which are qualitative and some quantitative},
  author  = {Lauritzen, Steffen L. and Wermuth, Nanny},
  journal = {Annals of Statistics},
  volume  = {17},
  number  = {1},
  pages   = {31--57},
  year    = {1989},
  doi     = {10.1214/aos/1176347003}
}

@article{sklar1959fonctions,
  title   = {Fonctions de r{\'e}partition {\`a} $n$ dimensions et leurs marges},
  author  = {Sklar, Abe},
  journal = {Publications de l'Institut de Statistique de l'Universit{\'e} de Paris},
  volume  = {8},
  pages   = {229--231},
  year    = {1959}
}

@article{bedford2002vines,
  title   = {Vines: {A} new graphical model for dependent random variables},
  author  = {Bedford, Tim and Cooke, Roger M.},
  journal = {Annals of Statistics},
  volume  = {30},
  number  = {4},
  pages   = {1031--1068},
  year    = {2002},
  doi     = {10.1214/aos/1031689016}
}

@article{aas2009pair,
  title   = {Pair-copula constructions of multiple dependence},
  author  = {Aas, Kjersti and Czado, Claudia and Frigessi, Arnoldo and Bakken, Henrik},
  journal = {Insurance: Mathematics and Economics},
  volume  = {44},
  number  = {2},
  pages   = {182--198},
  year    = {2009},
  doi     = {10.1016/j.insmatheco.2007.02.001}
}

@article{dissmann2013selecting,
  title   = {Selecting and estimating regular vine copulae and application to financial returns},
  author  = {Di{\ss}mann, Jeffrey and Brechmann, Eike Christian and Czado, Claudia and Kurowicka, Dorota},
  journal = {Computational Statistics \& Data Analysis},
  volume  = {59},
  pages   = {52--69},
  year    = {2013},
  doi     = {10.1016/j.csda.2012.08.010}
}

@book{czado2019analyzing,
  title     = {Analyzing Dependent Data with Vine Copulas},
  author    = {Czado, Claudia},
  year      = {2019},
  publisher = {Springer},
  address   = {Cham},
  doi       = {10.1007/978-3-030-13785-4}
}

@book{joe2014dependence,
  title     = {Dependence Modeling with Copulas},
  author    = {Joe, Harry},
  year      = {2014},
  publisher = {Chapman \& Hall/CRC},
  address   = {Boca Raton, FL}
}

@article{nagler2022rvinecopulib,
  title   = {\pkg{rvinecopulib}: {H}igh performance algorithms for vine copula modeling},
  author  = {Nagler, Thomas and Vatter, Thibault},
  journal = {Journal of Statistical Software},
  volume  = {102},
  number  = {5},
  pages   = {1--34},
  year    = {2022},
  doi     = {10.18637/jss.v102.i05}
}

@misc{pyvinecopulib2024,
  title  = {\pkg{pyvinecopulib}: {P}ython bindings for \pkg{vinecopulib}},
  author = {Nagler, Thomas and Vatter, Thibault},
  year   = {2024},
  note   = {Python package version 0.6},
  howpublished = {\url{https://github.com/vinecopulib/pyvinecopulib}}
}

@article{rosenblatt1952remarks,
  title   = {Remarks on a multivariate transformation},
  author  = {Rosenblatt, Murray},
  journal = {Annals of Mathematical Statistics},
  volume  = {23},
  number  = {3},
  pages   = {470--472},
  year    = {1952},
  doi     = {10.1214/aoms/1177729394}
}

@article{nagler2016evading,
  title   = {Evading the curse of dimensionality in nonparametric density estimation with simplified vine copulas},
  author  = {Nagler, Thomas and Czado, Claudia},
  journal = {Journal of Multivariate Analysis},
  volume  = {151},
  pages   = {69--89},
  year    = {2016},
  doi     = {10.1016/j.jmva.2016.07.003}
}

@article{clayton1978model,
  title   = {A model for association in bivariate life tables and its application in epidemiological studies of familial tendency in chronic disease incidence},
  author  = {Clayton, David G.},
  journal = {Biometrika},
  volume  = {65},
  number  = {1},
  pages   = {141--151},
  year    = {1978}
}

@article{gumbel1960bivariate,
  title   = {Bivariate exponential distributions},
  author  = {Gumbel, Emil J.},
  journal = {Journal of the American Statistical Association},
  volume  = {55},
  number  = {292},
  pages   = {698--707},
  year    = {1960}
}

@article{frank1979,
  title   = {On the simultaneous associativity of $F(x,y)$ and $x+y-F(x,y)$},
  author  = {Frank, Maurice J.},
  journal = {Aequationes Mathematicae},
  volume  = {19},
  number  = {1},
  pages   = {194--226},
  year    = {1979}
}

@article{mccullagh1980regression,
  title   = {Regression models for ordinal data},
  author  = {McCullagh, Peter},
  journal = {Journal of the Royal Statistical Society, Series B},
  volume  = {42},
  number  = {2},
  pages   = {109--142},
  year    = {1980}
}

@article{kahn1962topological,
  title   = {Topological sorting of large networks},
  author  = {Kahn, Arthur B.},
  journal = {Communications of the ACM},
  volume  = {5},
  number  = {11},
  pages   = {558--562},
  year    = {1962},
  doi     = {10.1145/368996.369025}
}

@article{morris2019using,
  title   = {Using simulation studies to evaluate statistical methods},
  author  = {Morris, Tim P. and White, Ian R. and Crowther, Michael J.},
  journal = {Statistics in Medicine},
  volume  = {38},
  number  = {11},
  pages   = {2074--2102},
  year    = {2019},
  doi     = {10.1002/sim.8086}
}

@book{vanderweele2015explanation,
  title     = {Explanation in Causal Inference: {M}ethods for Mediation and Interaction},
  author    = {VanderWeele, Tyler J.},
  year      = {2015},
  publisher = {Oxford University Press},
  address   = {New York}
}

@book{laan2018targeted,
  title     = {Targeted Learning in Data Science: {C}ausal Inference for Complex Longitudinal Studies},
  author    = {van der Laan, Mark J. and Rose, Sherri},
  year      = {2018},
  publisher = {Springer},
  address   = {New York}
}

@manual{rcore2024,
  title        = {R: {A} Language and Environment for Statistical Computing},
  author       = {{R Core Team}},
  organization = {R Foundation for Statistical Computing},
  address      = {Vienna, Austria},
  year         = {2024},
  url          = {https://www.R-project.org/}
}

@book{agresti2002categorical,
  title     = {Categorical Data Analysis},
  author    = {Agresti, Alan},
  year      = {2002},
  edition   = {2nd},
  publisher = {Wiley},
  address   = {Hoboken, NJ}
}

@article{wright2017ranger,
  title   = {\pkg{ranger}: A Fast Implementation of Random Forests for High Dimensional Data in \proglang{C++} and \proglang{R}},
  author  = {Wright, Marvin N. and Ziegler, Andreas},
  journal = {Journal of Statistical Software},
  volume  = {77},
  number  = {1},
  pages   = {1--17},
  year    = {2017},
  doi     = {10.18637/jss.v077.i01}
}

@article{burton2006design,
  title   = {The design of simulation studies in medical statistics},
  author  = {Burton, Andrea and Altman, Douglas G. and Royston, Patrick and Holder, Roger L.},
  journal = {Statistics in Medicine},
  volume  = {25},
  number  = {24},
  pages   = {4279--4292},
  year    = {2006},
  doi     = {10.1002/sim.2673}
}

@article{laubenbacher2024digital,
  title   = {Digital twins in medicine},
  author  = {Laubenbacher, Reinhard and Sluka, James P. and Glazier, James A.},
  journal = {Nature Computational Science},
  volume  = {4},
  pages   = {164--171},
  year    = {2024},
  doi     = {10.1038/s43588-024-00609-5}
}

@article{chickering2002optimal,
  title   = {Optimal structure identification with greedy search},
  author  = {Chickering, David Maxwell},
  journal = {Journal of Machine Learning Research},
  volume  = {3},
  pages   = {507--554},
  year    = {2002}
}

@techreport{verma1988causal,
  title       = {Causal networks: Semantics and expressiveness},
  author      = {Verma, Thomas S. and Pearl, Judea},
  institution = {UCLA Cognitive Systems Laboratory},
  year        = {1988},
  note        = {Technical Report R-65}
}

@article{scholkopf2021causal,
  title   = {Toward causal representation learning},
  author  = {Sch{\"o}lkopf, Bernhard and Locatello, Francesco and Bauer, Stefan and Ke, Nan Rosemary and Kalchbrenner, Nal and Goyal, Anirudh and Bengio, Yoshua},
  journal = {Proceedings of the IEEE},
  volume  = {109},
  number  = {5},
  pages   = {612--634},
  year    = {2021},
  doi     = {10.1109/JPROC.2021.3058954}
}

@article{breiman2001random,
  title   = {Random forests},
  author  = {Breiman, Leo},
  journal = {Machine Learning},
  volume  = {45},
  number  = {1},
  pages   = {5--32},
  year    = {2001},
  doi     = {10.1023/A:1010933404324}
}

@article{imai2010identification,
  title   = {Identification, inference and sensitivity analysis for causal mediation effects},
  author  = {Imai, Kosuke and Keele, Luke and Yamamoto, Teppei},
  journal = {Statistical Science},
  volume  = {25},
  number  = {1},
  pages   = {51--71},
  year    = {2010},
  doi     = {10.1214/10-STS321}
}

@article{imai2013identification,
  title   = {Identification and sensitivity analysis for multiple causal mechanisms: Revisiting evidence from framing experiments},
  author  = {Imai, Kosuke and Yamamoto, Teppei},
  journal = {Political Analysis},
  volume  = {21},
  number  = {2},
  pages   = {141--171},
  year    = {2013},
  doi     = {10.1093/pan/mps040}
}

@article{neyman1923,
  title   = {On the application of probability theory to agricultural experiments. {E}ssay on principles. {S}ection 9},
  author  = {Neyman, Jerzy},
  journal = {Statistical Science},
  volume  = {5},
  number  = {4},
  pages   = {465--472},
  year    = {1990},
  note    = {Translated from the Polish original (1923)}
}

@article{joe1993,
  title   = {Parametric families of multivariate distributions with given margins},
  author  = {Joe, Harry},
  journal = {Journal of Multivariate Analysis},
  volume  = {46},
  number  = {2},
  pages   = {262--282},
  year    = {1993},
  doi     = {10.1006/jmva.1993.1061}
}

\end{document}